\documentclass[%
pof,%
 amsmath,amssymb,
 reprint,%
onecolumn,
]{revtex4-1}
\usepackage{CJKutf8}
\usepackage{graphicx}
\usepackage{dcolumn}
\usepackage{bm}

\begin{document}

\preprint{AIP/123-QED}

\begin{CJK}{UTF8}{}
\CJKfamily{mj}
\title{Electric double layer of spherical pH-responsive polyelectrolyte brushes in an electrolyte solution: a strong stretching theory accounting for excluded volume interaction and mass action law}

\author{Jun-Sik Sin(신준식)}
\email{js.sin@ryongnamsan.edu.kp}
\affiliation{Natural Science Center, \textbf{Kim Il Sung} University, Taesong District, Pyongyang, Democratic People's Republic of Korea}
\author{Il-Chon Choe(최일천)}
\affiliation{Natural Science Center, \textbf{Kim Il Sung} University, Taesong District, Pyongyang, Democratic People's Republic of Korea}
\author{Chol-Song Im(임철송)}
\affiliation{Faculty of Koryo medicine, Pyongyang University of Medical Sciences, Pyongyang, Democratic People's Republic of Korea}


\begin{abstract}
\large
 
In this paper, we study the electrostatics of pH-responsive polyelectrolyte-grafted spherical particles by using a strong stretching theory that takes into account the excluded volume interaction and the density of chargeable sites on the polyelectrolyte molecules. 

Based on the free energy formalism, we obtain self-consistent field equations for determining the structure and electrostatics of spherical polyelectrolyte brushes. 

First, we find that the smaller the radius of the inner core, the longer the height of the polyelectrolyte brush. Then, we also prove that an increase in excluded volume interaction yields an swelling of the polyelectrolyte brush height. In addition, we demonstrate how the effect of pH, bulk ionic concentration, and lateral separation between adjacent polyelectrolyte chains on the electrostatic properties of a spherical polyelectrolyte brush is affected by the radius of inner core, the excluded volume interaction and the chargeable site density.

\end{abstract}
\maketitle
\end{CJK}

\large
\section{\label{sec:level1}Introduction}

When polyelectrolyte chains are densely grafted onto a spherical particle, the interaction between the polyelectrolyte chain molecules causes each polymer chain to orient nearly perpendicular to the surface, resulting in the formation of a spherical polyelectrolyte brush. \cite%
{mura_natmat_2013, miller_japs_2014, wang_biomol_2013, zhu_nanotech_2012}
While planar polyelectrolyte brushes are mainly formed in nanochannels and exhibit various functions such as current rectifiers \cite%
{groot_acsapp_2013, ali_jacs_2009, vilozny_nanoscale_2013, ali_acsnano_2009}, ion sensing \cite%
{yammen_jacs_2008,ali_macromol_2010, umehara_pnas_2009}, flow valves \cite%
{ali_jacs_2010,moya_acid_2005, xin_csr_2010}
 and solute transport \cite%
{ohshima_pof_2022}
, spherical polyelectrolyte brushes have been used in various fields such as drug delivery \cite%
{mura_natmat_2013, wang_biomol_2013}, oil recovery \cite%
{miller_japs_2014} or emulsion stabilization \cite%
{zhu_nanotech_2012, heuzey_pof_2019}.

Micro/nanoparticles have been designed and used to release drugs sensitively under local heterogeneous pH changes to deliver specific drugs to damaged cells and tissues \cite%
{vafai_pof_2021}, and for applications such as oil recovery, emulsion stabilization \cite%
{ vannozzi_pof_2019}, water recovery, etc.

The growing demand for pH-responsive spherical polymer brushes requires additional studies of the electrostatics of such spherical polyelectrolyte brushes. Studies on spherical polyelectrolyte brush were mostly based on scaling theory and rigorous self-consistent field model. \cite%
{pincus_macromol_1991,ross_macromol_1992,borisov_jp_1991,joanny_macromol_1993,borisov_macromol_1994, zhulina_macromol_1995_1,zhulina_macromol_1995_2,zhulina_sm_2012,zhulina_macromol_1994,misra_macromol_1989,das_jpcb_2015,das_jap_2015,das_rscadv_2015,
das_jpcb_2016}

However, such models assume that the hydrogen ion distribution inside and outside the polyelectrolyte layer follows the Boltzmann distribution. Das confirmed that the description of such hydrogen ion concentration on a nanochannel grafted with pH-responsive polyelectrolyte brushes indeed violates thermodynamics and proposed to accept a description of hydrogen ion concentration that properly takes into account the ionization of the polyelectrolyte layer. 
In such a formalism, based on the free energy description of an electrical double layer with explicit consideration of polyelectrolyte charge and hydrogen ion concentration, the hydrogen ion concentration profile is derived by minimizing the free energy with respect to hydrogen ion concentration rather than the Boltzmann distribution. \cite%
{das_jpcb_2017,das_csb_2016}
These methods have been extended to the case where pH-responsive polyelectrolyte brushes take into account ion size effects and to the case of spherical polyelectrolyte brushes.
However, such a formalism assumes that the polyelectrolyte brush is described only by a balance of elastic energy and excluded volume effects.  Furthermore, they considered the extremely simple case where the length of all chains of the polyelectrolyte brush is uniform.

Recently, Das's group \cite%
{das_sm_2019_1, das_pre_2020, das_sm_2019_2, das_pof_2020}
 proposed a more advanced version of strong stretching, a self-consistent field theory that takes into account excluded volume interaction and generic mass action law, to further develop the theory of polyelectrolyte brushes \cite%
{zhulina_jcp_1997, borisov_jcp_2017, zhulina_macromol_2000}, and to apply various transport and interfacial phenomena to obtain good results. However, the models are related to the planar polyelectrolyte brushes, but not to the spherical polyelectrolyte brush.

Very recently, the author of \cite%
{sin_jcp_2022} improved the strong stretching theory by taking into account Born energy and ionic size, and investigated the osmotic pressure between two pH-responsive polyelectrolyte brushes and the height of polyelectrolyte brushes.

Although Budkov's group \cite%
{bukov_jpcm_2018, bukov_fpe_2019, bukov_pccp_2020, bukov_jpcm_2020, bukov_jml_2021}
addressed probability density functions for star-like molecules or zwitterionic and multipolar molecules by using nonlocal statistical field theory, 
they did not investigate spherical polyelectrolyte brushes.

To fill the gap, we extend the theory of those papers to a system of spherical polyelectrolyte brushes to obtain novel properties in such a system.
We first assume that the electrostatic effects of the polyelectrolyte brush constitute the free energy of the system, taking into account the elastic and excluded volume effects and the ionization energy.
Then, based on the variational principle, we determine the self-consistent field equations to determine the electric field distribution and ion distribution around the spherical polyelectrolyte brushes, the monomer distribution of the polyelectrolyte chain and the polyelectrolyte chain end distribution.
Finally, we provide results on the electrical double-layer electrostatic potential and polyelectrolyte monomer concentration of spherical polyelectrolyte brushes with various physical quantities such as radius, pH, bulk ionic concentration, lateral separaion between adjacent polyelectrolyte chain molecules to determine the varying nature of the pH response.

\section{Theory}

We consider a pH-responsive, spheriacal polyelectrolyte brush with the radius $r_0$ of the inner core.
The spherical polyelectrolyte brushe has an uncharged metallic nanoscale core, which cannot pass through both the ions and the solvent.
The paper is applicable to more general cases without the need for severe constraints such as constancy of the polyelectrolyte chain length.
It is only required that the grafted density of polyelectrolytes is large enough to allow strong stretching.
\begin{figure}
\begin{center}
\includegraphics[width=0.5\textwidth]{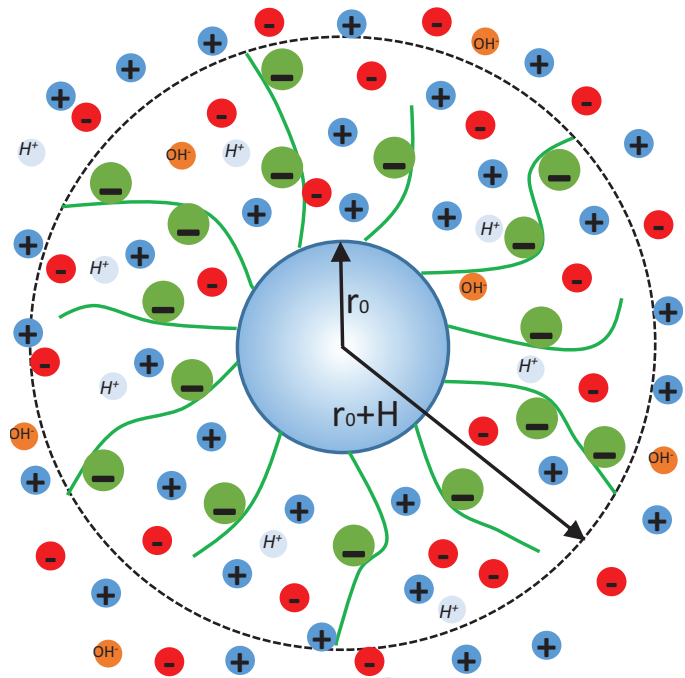}
\caption{(Color online) Schematic of a spherical polyelectrolyte brush in an electrolyte solution. A spherical polyelectrolyte brush consists of a inner core and a dense polyelectrolyte layer, where every polyelectrolyte chain is strongly stretched toward radical direction. Here the polyelectrolyte ions are represented in green, electrolytic anions and cations are represented in red and blue, respectively. Hydrogen ions and hydroxyl ions are represented in cyan and orange.
}
\label{fig:1}
\end{center}
\end{figure}

As in \cite%
{das_sm_2019_1},  taking into account the effects of the polyelectrolyte excluded volume interactions and mass action law, we apply the strong stretching theory to a spherical nanoparticle. 

The theory presents the equilibrium structure of the polyelectrolyte brushes and the equilibrium electrostatics of the brush-induced electric double layer by minimizing the total free energy of a given brush that consists of the elastic, excluded volume, electrostatic, and ionization energies of the brush and the electrostatic energy of the brush-induced electric double layer. 

We determine the equlibrium structrue by minimizing the total energy functional (F) of the polyelectrolyte brush system. $F$ consists of the elastic ($F_{els}$), excluded volume ($F_{EV}$), electrostatic ($F_{elec}$), and ionization ($F_{ion}$) free energies of a polyelectrolye brush molecule and the electrostatic energy of the electric double layer($F_{EDL}$) induced by this brush. 
\begin{equation}
 F = F_{els}  + F_{EV}  + F_{elec}  + F_{ion}  + F_{EDL} 
\label{eqn:1}
\end{equation}
In the same way as Das et al \cite%
{das_sm_2019_1}, we can express $F_{els}$ as
\begin{equation}
\frac{{F_{els} }}{{k_B T}} = \frac{3}{{2pa^2 r_0 ^2 }}\int_{r_0 }^{r_0  + H} {g\left( {r'} \right)r'^2 dr'} \int_{r_0 }^{r'} {E\left( {r,r'} \right)dr} 
\label{eqn:2}
\end{equation}
\begin{equation}
\frac{{F_{EV} }}{{k_B T}} = \frac{1}{{\sigma a^3 r_0 ^2 }}\int_{r_0 }^{r_0  + H} {f_{EV} \left[ {\phi \left( r \right)} \right]r^2 dr} 
\label{eqn:3}
\end{equation}
where  $k_B T, p, a, H$  and $\sigma\propto\frac{1}{{l^2 }}$ ($l$ is the lateral separation between the adjacent polyelectrolyte brushes), respectively, represent the thermal energy, chain rigidity, Kuhn length, brush height, and polyelectrolyte brush grafting density. Also $\phi\left(r\right)$ and $f_{EV} \left[ {\phi \left( r \right)} \right]\left( { \approx v\phi ^2  + \omega \phi ^3 } \right)$, respectively, denote the dimensionless monomer distribution profile of a polyelectrolyte chain and the nondimensionalized per unit volume free energy associated with the excluded volume interactions. Also, $E\left({r, r'}\right) = \frac{{dr}}{{dn}}$ expresses chain stretching (for a chain whose end is at $r'$) at a location $r$ and $g\left(r'\right)$(the normalized chain end distribution function) is expressed as
\begin{equation}
\frac{1}{{r_0 ^2 }}\int_{r_0 }^{r_0  + H} {g\left( {r'} \right)r'^2 dr'}  = 1
\label{eqn:4}
\end{equation}
Next, $F_{elec}  + F_{EDL}$ can be expressed as 
\begin{eqnarray}
 \frac{F_{elec}  + F_{EDL}}{k_B T}= \frac{1}{{\sigma k_B Tr_0 ^2 }}\int_{r_0 }^\infty  {\left[ { - \frac{{\varepsilon _0 \varepsilon _r }}{2}\left| {\frac{{d\psi }}{{dr}}} \right|^2  + e\psi \left( {n_ +   - n_ -   + n_{H^ +  }  - n_{OH^ -  } } \right)} \right]r^2 dr}  \nonumber \\ 
  - \frac{1}{{\sigma k_B Tr_0 ^2 }}\int_{r_0 }^{r_0  + H} {\left[ {e\psi n_{A^ -  } \phi } \right]r^2 dr}  \nonumber \\ 
  + \frac{1}{{\sigma r_0 ^2 }}\int_{r_0 }^\infty  {\left\{ {n_ +  \left[ {\ln \left( {\frac{{n_ +  }}{{n_{ + ,\infty } }}} \right) - 1} \right] + n_ -  \left[ {\ln \left( {\frac{{n_ -  }}{{n_{ - ,\infty } }}} \right) - 1} \right]} \right\}} r^2 dr \nonumber\\ 
  + \frac{1}{{\sigma r_0 ^2 }}\int_{r_0 }^\infty  {\left\{ {n_{H^ +  } \left[ {\ln \left( {\frac{{n_{H^ +  } }}{{n_{H^ +  ,\infty } }}} \right) - 1} \right] + n_{OH^ +  } \left[ {\ln \left( {\frac{{n_{OH^ +  } }}{{n_{OH^ +  ,\infty } }}} \right) - 1} \right]} \right\}r^2 dr}  \nonumber \\ 
  + \frac{1}{{\sigma r_0 ^2 }}\int_{r_0 }^\infty  {\left( {n_{ + ,\infty }  + n_{ - ,\infty }  + n_{H^ +  ,\infty }  + n_{OH^ -  ,\infty } } \right)r^2 dr}
\label{eqn:5}
\end{eqnarray}
where $\psi$  is the electrostatic potential, $n_i$ and $n_{i,\infty }$ are the number density and bulk number density for ion $i$ (where $i =  \pm ,H^ +  ,OH^ - $). Here $n_{i,\infty }  = 10^3 N_A c_{i,\infty }$, $c_{i,\infty}$ and $N_A$ are the bulk ionic concentration, and Avogardro number, respectively. $e$ is the elementary charge, $\varepsilon _0$ and $\varepsilon _r$ and  are the permittivity of vaccum and relative permittivity of the solution, respectively.
As in \cite%
{das_sm_2019_1}, a pH-responsive polyelectrolyte brush is charged by an acidlike dissociation of $HA$ producing $H^+$  and $A^ -$  ions. $K_a$ is the ionization constant for the acidlike dissociation process. The number density of these ions 
$A^ -$  is expressed as $n_{A^ -  }$.  
Consequently,  $F_{ion}$ can be written as
 \begin{eqnarray}
\frac{{F_{ion} }}{{k_B T}} = \frac{1}{{\sigma a^3 r_0 ^2 }}\int_{r_0 }^{r_0  + H} {\phi \left[ {\left( {1 - \frac{{n_{A^ -  } }}{\gamma }} \right)\ln \left( {1 - \frac{{n_{A^ -  } }}{\gamma }} \right) + \frac{{n_{A^ -  } }}{\gamma } \ln \left( {\frac{{n_{A^ -  } }}{\gamma }} \right) + \frac{{n_{A^ -  } }}{\gamma }\ln \left( {\frac{{n_{H^ +  ,\infty } }}{{{K_a}' }}} \right)} \right]r^2 dr} 
\label{eqn:6}
\end{eqnarray}
where ${K_a}'  = 10^3 N_A K_a ,n_{H^ +  ,\infty }  = 10^3 N_A c_{H^ +  ,\infty }$  is the bulk concentration of the ions, which can be related with bulk $pH$ or $pH_\infty$ as $ c_{H^ +  ,\infty }  = 10^{ - pH_\infty}$, and $\gamma$ is the density of polyelectrolyte chargeable sites.
Substituting Eq. (\ref{eqn:2}), Eq. (\ref{eqn:3}), Eq. (\ref{eqn:5}) and Eq. (\ref{eqn:6}) into Eq. (\ref{eqn:1}), we obtain the full expression for the free energy as
 \begin{eqnarray}
 \frac{F}{{k_B T}} = \frac{3}{{2pa^2 r_0 ^2 }}\int_{r_0 }^{r_0  + H} {g\left( {r'} \right)r'^2 dr'} \int_0^{r'} {E\left( {r,r'} \right)dr}  + \frac{1}{{\sigma a^3 r_0 ^2 }}\int_{r_0 }^{r_0  + H} {f_{EV} \left[ {\phi \left( r \right)} \right]r^2 dr} \nonumber\\
+ \frac{1}{{\sigma k_B Tr_0 ^2 }}\int_{r_0 }^\infty  {\left[ { - \frac{{\varepsilon _0 \varepsilon _r }}{2}\left| {\frac{{d\psi }}{{dr}}} \right|^2  + e\psi \left( {n_ +   - n_ -   + n_{H^ +  }  - n_{OH^ -  } } \right)} \right]r^2 dr} \nonumber\\ 
- \frac{1}{{\sigma k_B Tr_0 ^2 }}\int_{r_0 }^{r_0  + H} {\left[ {e\psi n_{A^ -  } \phi } \right]r^2 dr} \nonumber \\ 
  + \frac{1}{{\sigma r_0 ^2 }}\int_{r_0 }^\infty  {\left\{ {n_ +  \left[ {\ln \left( {\frac{{n_ +  }}{{n_{ + ,\infty } }}} \right) - 1} \right] + n_ -  \left[ {\ln \left( {\frac{{n_ -  }}{{n_{ - ,\infty } }}} \right) - 1} \right] + n_{H^ +  } \left[ {\ln \left( {\frac{{n_{H^ +  } }}{{n_{H^ +  ,\infty } }}} \right) - 1} \right]} \right\}r^2 dr}  \nonumber\\ 
  + \frac{1}{{\sigma r_0 ^2 }}\int_{r_0 }^\infty  {\left\{ {n_{OH^ +  } \left[ {\ln \left( {\frac{{n_{OH^ +  } }}{{n_{OH^ +  ,\infty } }}} \right) - 1} \right] + \left( {n_{ + ,\infty }  + n_{ - ,\infty }  + n_{H^ +  ,\infty }  + n_{OH^ -  ,\infty } } \right)} \right\}r^2 dr} \nonumber \\ 
  + \frac{1}{{\sigma a^3 r_0 ^2 }}\int_{r_0 }^\infty  {\phi \left[ {\left( {1 - \frac{{n_{A^ -  } }}{\gamma }} \right)\ln \left( {1 - \frac{{n_{A^ -  } }}{\gamma }} \right) + \frac{{n_{A^ -  } }}{\gamma }\ln \left( {\frac{{n_{A^ -  } }}{\gamma }} \right) + \frac{{n_{A^ -  } }}{\gamma }\ln \left( {\frac{{n_{H^ +  ,\infty } }}{{{K_a}' }}} \right)} \right]r^2 dr}
\label{eqn:7}
\end{eqnarray}

Eq. (\ref{eqn:7}) should be minimized according to the variational principle in the presence of the following constraints:
 \begin{equation}
N = \int_{r_0 }^{r'} {\frac{{dr}}{{E\left( {r,r'} \right)}}} 
\label{eqn:8}
\end{equation}
\begin{equation}
N = \frac{1}{{\sigma a^3 r_0 ^2 }}\int_{r_0 }^{r_0  + H} {\phi \left( r \right)r^2 dr},
\label{eqn:9}
\end{equation}
where $N$ denotes the number of Kuhn monomer in a polyelectrolyte brush molecule.
Moreover, we have $\phi \left( r \right)$ related to the functions $g$ and $E$ as
 \begin{equation}
\phi \left( r \right) = \frac{{\sigma a^3 }}{{r_0 ^2 }}\int_r^{r_0  + H} {\frac{{g\left( {r'} \right)r'^2 dr'}}{{E\left( {r,r'} \right)}}} 
\label{eqn:10}
\end{equation}
This minimization process provides the self-consistent equations at the equilibrium state of the system. In the present study, we have considered fully flexible polyelectrolyte brush chains ($p=1$) only.
These equations are provided as follows:
 \begin{equation}
n_{A^ -  }  = \frac{{{K_a}' \gamma }}{{{K_a}'  + n_{H^ +  ,\infty } \exp \left( { - \gamma a^3 \frac{{e\psi }}{{k_B T}}} \right)}}
\label{eqn:11}
\end{equation}
\begin{equation}
\varepsilon _0 \varepsilon _r \left( {\frac{{d^2 \psi }}{{dr^2 }} + \frac{2}{r}\frac{{d\psi }}{{dr}}} \right) + e\left( {n_ +   - n_ -   + n_{H^ +  }  - n_{OH^ -  }  - n_{A^ -  } \phi } \right) = 0, r_0  \le r \le r_0  + H 
\label{eqn:12}
\end{equation}
\begin{equation}
\varepsilon _0 \varepsilon _r \left( {\frac{{d^2 \psi }}{{dr^2 }} + \frac{2}{r}\frac{{d\psi }}{{dr}}} \right) + e\left( {n_ +   - n_ -   + n_{H^ +  }  - n_{OH^ -  } } \right) = 0, r_0  + H \le r < \infty
\label{eqn:13}
\end{equation}
\begin{equation}
n_ \pm   = n_{ \pm ,\infty } \exp \left( { \mp \frac{{e\psi }}{{k_B T}}} \right),
\label{eqn:14}
\end{equation}
\begin{equation}
n_{H^ +  }  = n_{H^ +  ,\infty } \exp \left( { - \frac{{e\psi }}{{k_B T}}} \right),
\label{eqn:15}
\end{equation}
\begin{equation}
n_{OH^ -  }  = n_{OH^ -  ,\infty } \exp \left( {\frac{{e\psi }}{{k_B T}}} \right),
\label{eqn:16}
\end{equation}
\begin{equation}
\phi \left( r \right) = \frac{v}{{3\omega }}\left( {\left[ {1 + \kappa ^2 f\left( r \right)} \right]^{1/2}  - 1} \right),
\label{eqn:17}
\end{equation}
where $f\left( r \right) = \lambda  - \frac{{r_0 ^2 \left( {r - r_0 } \right)^2 }}{{r^2 }} + \beta \frac{{{K_a}' \gamma }}{{{K_a}'  + n_{H^ +  ,\infty } \exp \left( { - \gamma a^3 \frac{{e\psi }}{{k_B T}}} \right)}}\psi \\
 - \rho \left( {1 - \frac{{{K_a}' }}{{{K_a}'  + n_{H^ +  ,\infty } \exp \left( { - \gamma a^3 \frac{{e\psi }}{{k_B T}}} \right)}}} \right)\ln \left( {1 - \frac{{{K_a}' }}{{{K_a}'  + n_{H^ +  ,\infty } \exp \left( { - \gamma a^3 \frac{{e\psi }}{{k_B T}}} \right)}}} \right) \\
- \rho \frac{{{K_a}' }}{{{K_a}'  + n_{H^ +  ,\infty } \exp \left( { - \gamma a^3 \frac{{e\psi }}{{k_B T}}} \right)}}\ln \frac{{{K_a}' }}{{{K_a}'  + n_{H^ +  ,\infty } \exp \left( { - \gamma a^3 \frac{{e\psi }}{{k_B T}}} \right)}} - \rho \frac{{{K_a}' }}{{{K_a}'  + n_{H^ +  ,\infty } \exp \left( { - \gamma a^3 \frac{{e\psi }}{{k_B T}}} \right)}}\ln \left( {\frac{{n_{H^ +  ,\infty } }}{{{K_a}' }}} \right)$.
\begin{equation}
E\left( {r,r'} \right) = \frac{\pi }{{2N}}\sqrt {r'^2  - r^2 }
\label{eqn:18}
\end{equation}
\begin{equation}
\left( {q_{tot} } \right)_{H}  = 0
\label{eqn:19}
\end{equation}
\begin{equation}
\left( {q_{tot} } \right)_{H}  = \frac{e}{{\sigma r_0 ^2 }}\int_{r_0 }^\infty  {\left( {n_ +   - n_ -   + n_{H^ +  }  - n_{OH^ -  }  - \phi n_{A^ -  } } \right)r^2 dr}
\label{eqn:20}
\end{equation}
\begin{equation}
g\left( r \right) = \frac{r}{{\sigma Na^3 }}\left[ {\frac{{\phi \left( H \right)}}{{\sqrt {H^2  - r^2 } }} - \int_r^{r_0  + H} {\frac{{d\phi \left( {r'} \right)}}{{dr'}}\frac{{dr'}}{{\sqrt {r'^2  - r^2 } }}} } \right]
\label{eqn:21}
\end{equation}
Eq. (\ref{eqn:11}) expresses the expanded form of the mass action law. Eq. (\ref{eqn:12}) and  Eq. (\ref{eqn:13})  describes the electrostatic potential distribution both inside (\(r_0  \le r \le r_0  + H\)) and outside ( \(r_0  + H \le r\)) the brushes, respectively.

Eq. (\ref{eqn:14})-(\ref{eqn:16}) provide the ion number densities which follow Boltzmann distribution. Eq. (\ref{eqn:17}) provides the monomer distribution profile using virial coefficients  $\nu$ and $w$, parameters  \(\rho  = \frac{{8a^2 N^2 }}{{3\pi ^2 }}\),  (\(\lambda  =  - \lambda _1 \rho  =  - \lambda _1 \frac{{8a^2 N^2 }}{{3\pi ^2 }}\)
, $\lambda_1$ is the Lagrange multiplier yielded by the constraint expressed in \ref{eqn:9}. \(\beta  = \frac{{8N^2 ea^5 }}{{3\pi ^2 k_B T}}\).

Eq. (\ref{eqn:18}) provides the local stretching of the polyelectrolyte brush. Eq. (\ref{eqn:18}) is used as the condition for obtaining the equilibrium brush height $H$.

Eq. (\ref{eqn:19}) and Eq. (\ref{eqn:20}) represents  electric charge neutrality inside electrolyte solution containing spherical pH-responsive polyelectrolyte brushes.

Finally, Eq. (\ref{eqn:21}) expresses the normalized distribution $g\left(r\right)$ for the end of the polyelectrolyte brush end (obtained from the condition $\int_{r_0 }^{r_0  + H} {g\left( {r'} \right)dr'}  = 1$  . 

The configuration and the electrostatics of induced electric double layer of the polyelectrolyte brush molecules is obtained by solving Eq. (\ref{eqn:11}) - (\ref{eqn:20}). 
Of course, the electrostatic potential and the brush height for a spherical pH-responsive polyelectrolyte brush in a good solvent is self-consistently solved by using  Eq. (\ref{eqn:11}), (\ref{eqn:13}) - (\ref{eqn:16}) and considering the following boundary condition:
\begin{eqnarray}
 \left( \psi  \right)_{r = \left( {r_0  + H} \right)^ -  }  = \left( \psi  \right)_{r = \left( {r_0  + H} \right)^ +  } , \nonumber \\ 
 \left( {\frac{{d\psi }}{{dr}}} \right)_{r = \left( {r_0  + H} \right)^ -  }  = \left( {\frac{{d\psi }}{{dr}}} \right)_{r = \left( {r_0  + H} \right)^ +  } , \nonumber \\ 
 \left( \psi  \right)_{r = \infty }  = 0, \\ 
 \left( {\frac{{d\psi }}{{dr}}} \right)_{r = r_0 }  = 0.\nonumber
 \end{eqnarray}
The solution provides $\phi ,\psi ,g\left( x \right), H, n_{A^ -  } ,n_i \left( {i =  \pm ,H^ +  ,OH^ -  } \right)$ and therefore provides the complete equilibrium description of the system.

\section{Results and Discussion}

  We account for the influence of the excluded volume interaction and expanded form of mass action law on pH-responsive brushes to be in a 'good' solvent.
In all the subsequent calculations, the temperature and the monomer number on a polyelectrolyte chain is are $298K$ and 200, respectively. In addition, Kuhn length  is $a=1nm$
\begin{figure}
\centering
\includegraphics[width=0.6\textwidth]{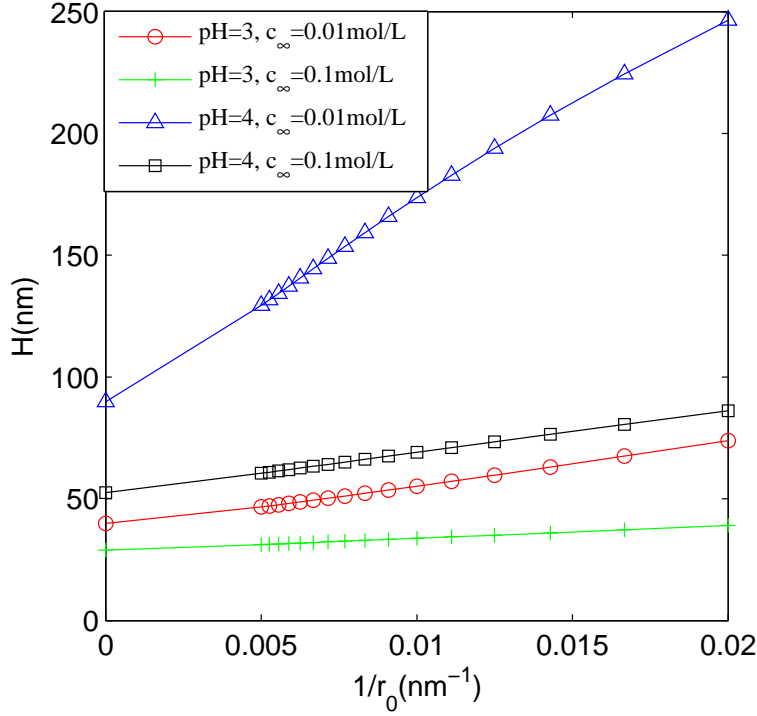}
\caption{(Color online) Variation of equilibrium brush height $H$ with curvature for different $pH$  and $c_\infty$. $pH=3, 4$, $c_\infty=0.1mol/L, 0.01mol/L, \gamma a^3=1,  l_0=10nm$.}
\label{fig:2}
\end{figure}
\begin{figure}
\includegraphics[width=1\textwidth]{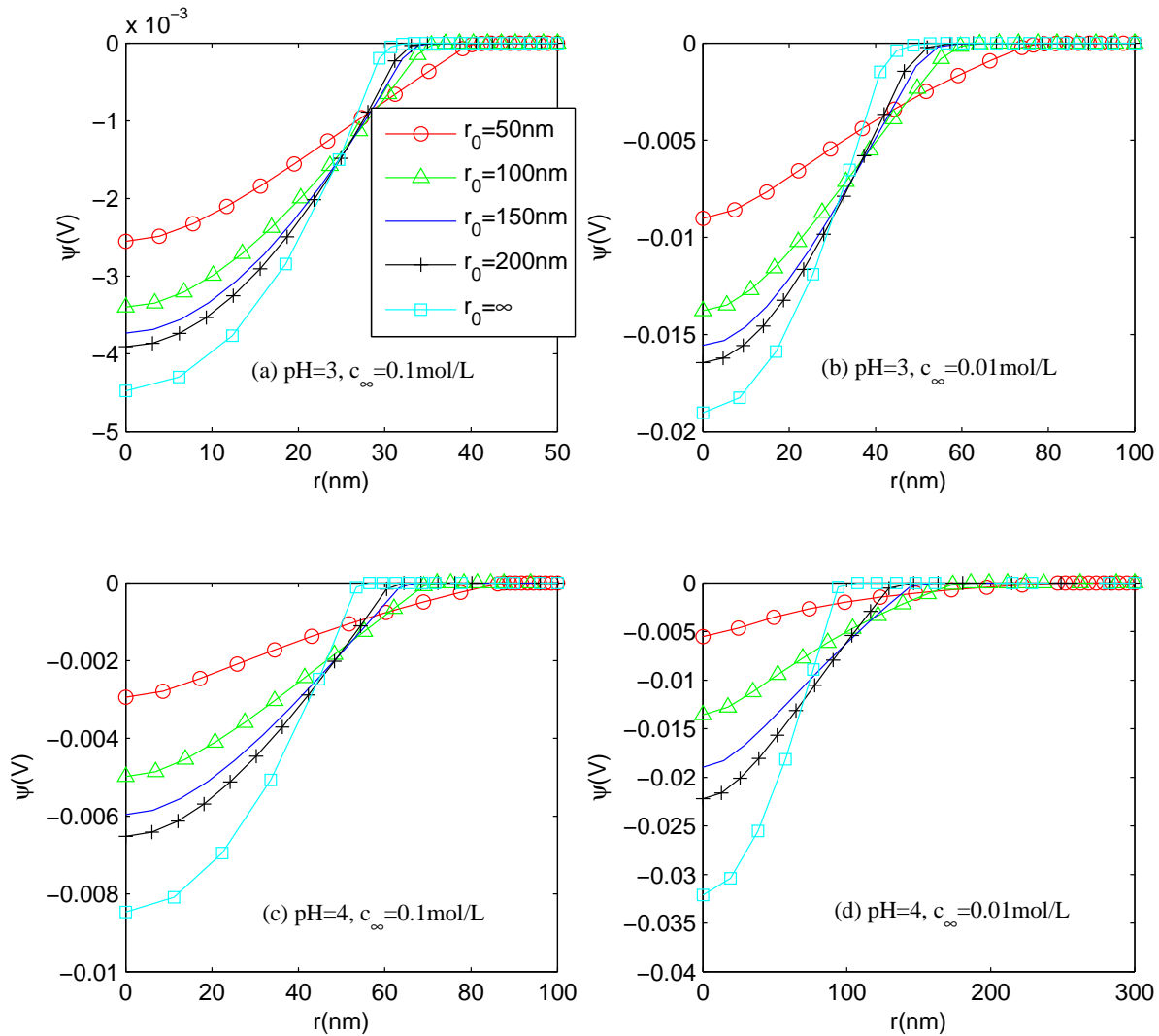}
\caption{(Color online) Electrostatic potential as a function of the distance from the inner core surface for (a)  $pH=3$, $c_\infty =0.1mol/L$; (b) $pH=3$, $c_\infty =0.01mol/L$; (c) $pH=4$, $c_\infty =0.1mol/L$; (d) $pH=4$, $c_\infty =0.01mol/L$ at radius $r_0= 50nm,100nm,150nm, 200nm$. Other parameters are the same as in Fig. \ref{fig:2} }
\label{fig:3}
\end{figure}
\begin{figure}
\includegraphics[width=1\textwidth]{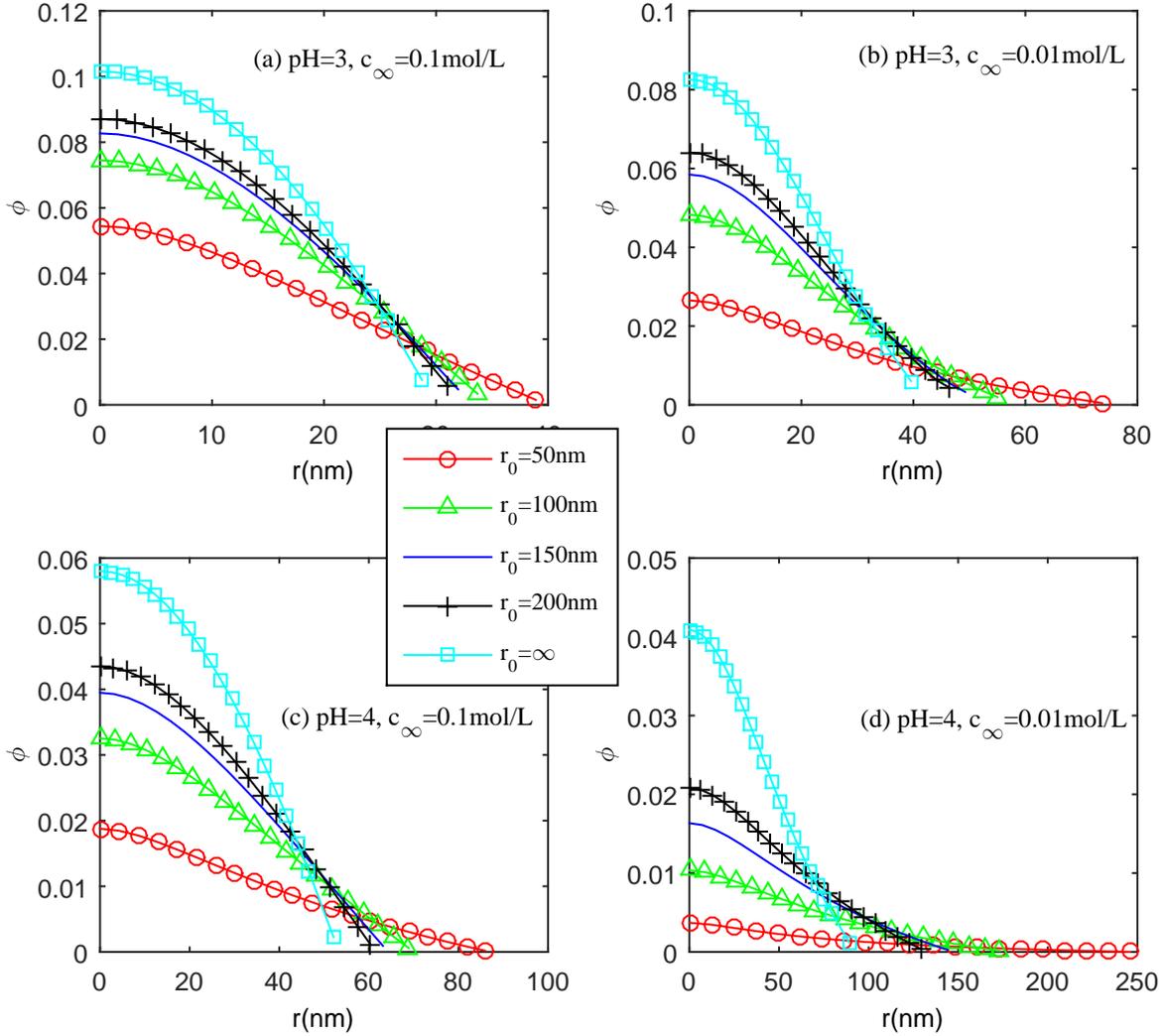}
\caption{(Color online) Monomer distribution profiles for (a)  $pH=3$, $c_\infty =0.1mol/L$; (b) $pH=3$, $c_\infty =0.01mol/L$; (c) $pH=4$, $c_\infty =0.1mol/L$; (d) $pH=4$, $c_\infty =0.01mol/L$ at radius of the inner core $r_0=50nm,100nm,150nm, 200nm$. Other parameters are the same as in Fig. \ref{fig:3}}
\label{fig:4}
\end{figure}
 \begin{figure}
\includegraphics[width=1\textwidth]{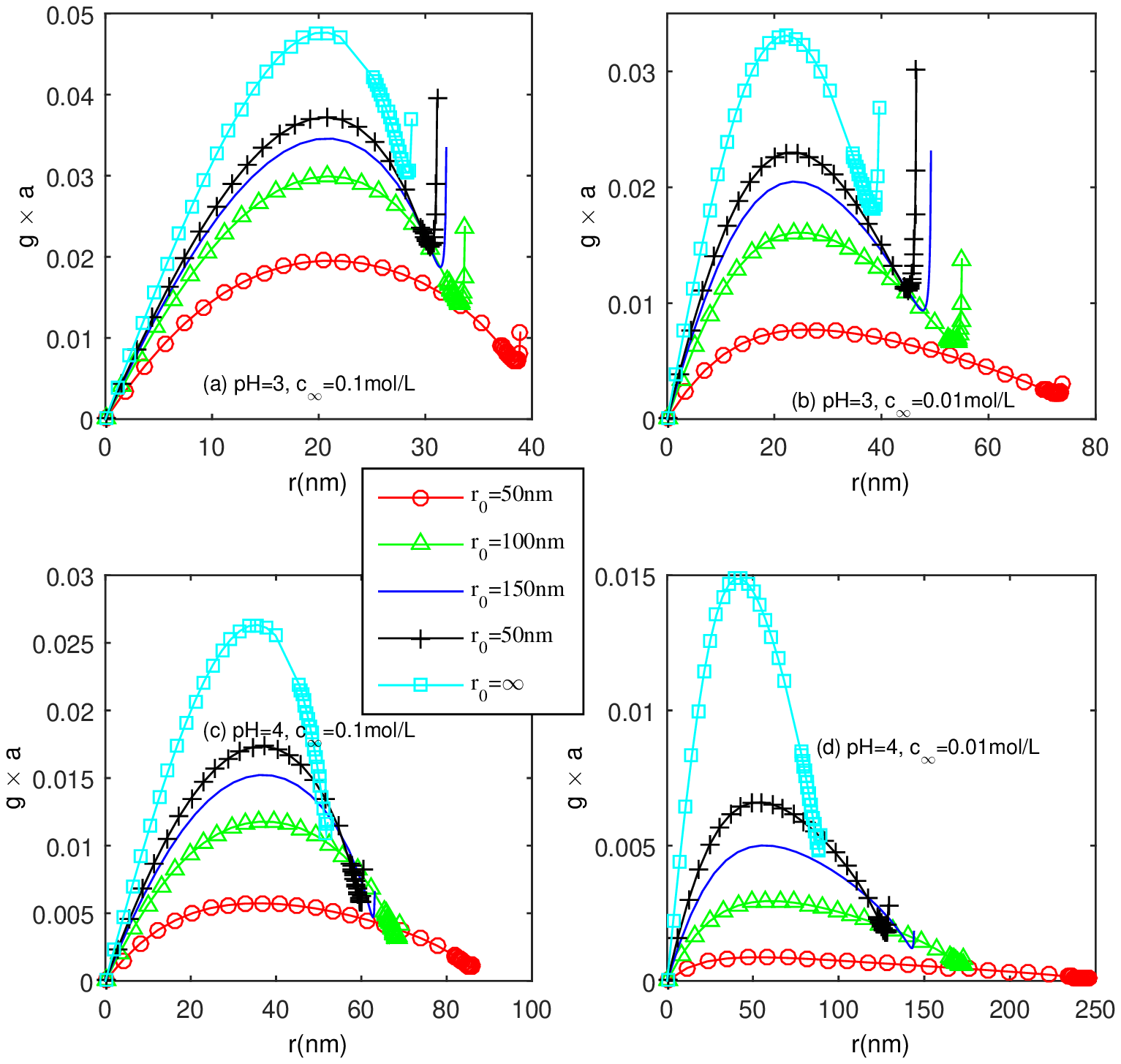}
\caption{(Color online) Chain end distribution profiles for (a)  $pH=3$, $c_\infty =0.1mol/L$; (b) $pH=3$, $c_\infty =0.01mol/L$; (c) $pH=4$, $c_\infty =0.1mol/L$; (d) $pH=4$, $c_\infty =0.01mol/L$ at radius of the inner core $r_0=50nm,100nm,150nm, 200nm$. Other parameters are the same as in Fig. \ref{fig:3}}
\label{fig:5}
\end{figure}
\begin{figure}
\includegraphics[width=\linewidth]{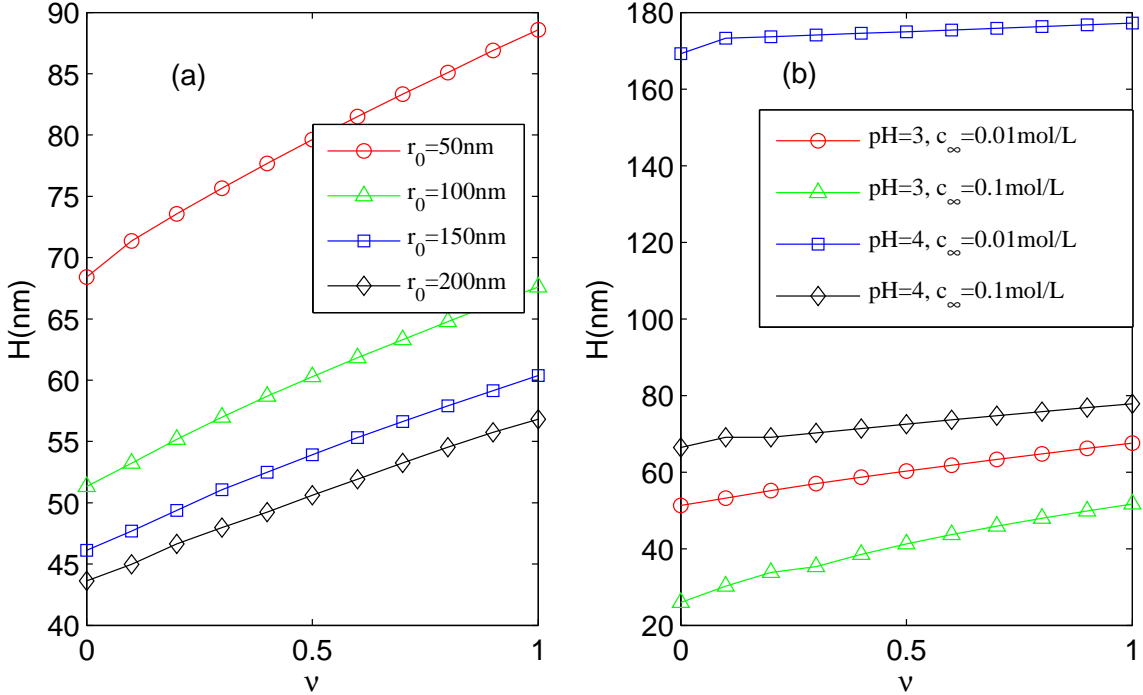}
\caption{(Color online) Variation of equilibrium brush height H with excluded volume interaction factor $\nu$ (a) for $pH=3, c_\infty=0.01mol/L$ at $r_0=50nm,100nm,150nm, 200nm$; (b) $pH=3, 4$; $c_\infty=0.01mol/L, 0.1mol/L$, $r_0=100nm$. Other parameters are the same as in Fig. \ref{fig:2}}
\label{fig:6}
\end{figure}
\begin{figure}
\includegraphics[width=0.65\textwidth]{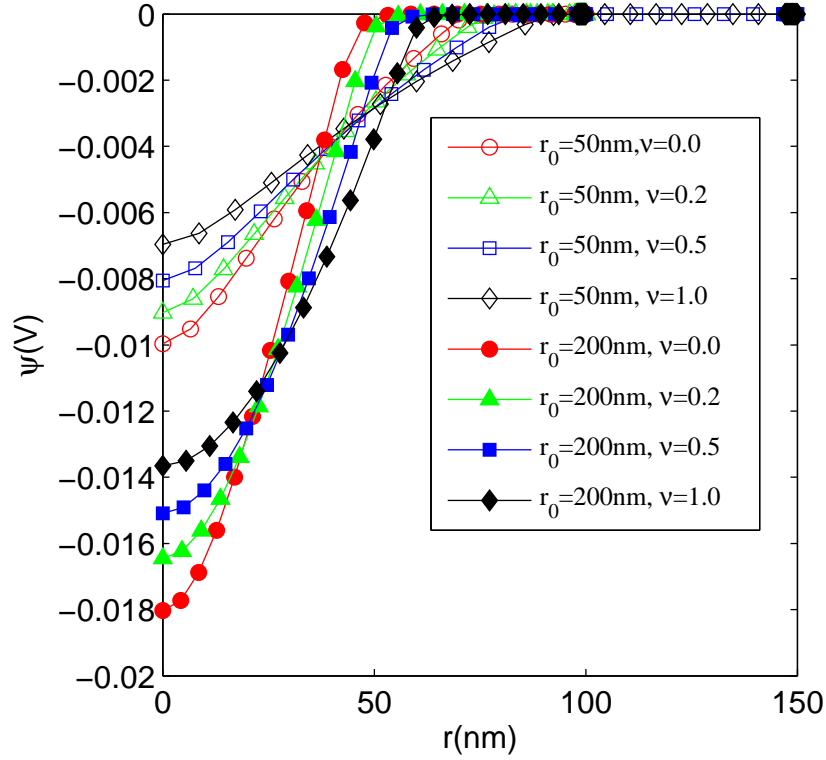}
\caption{(Color online) Electrostatic potential profiles for $\nu =0.0,0.2,0.5,1.0$ at $r_0=50nm, 200nm$. $pH=3, cp=0.01mol/L$. Other parameters are the same as in Fig. \ref{fig:6}}
\label{fig:7}
\end{figure}
\begin{figure}
\includegraphics[width=0.65\textwidth]{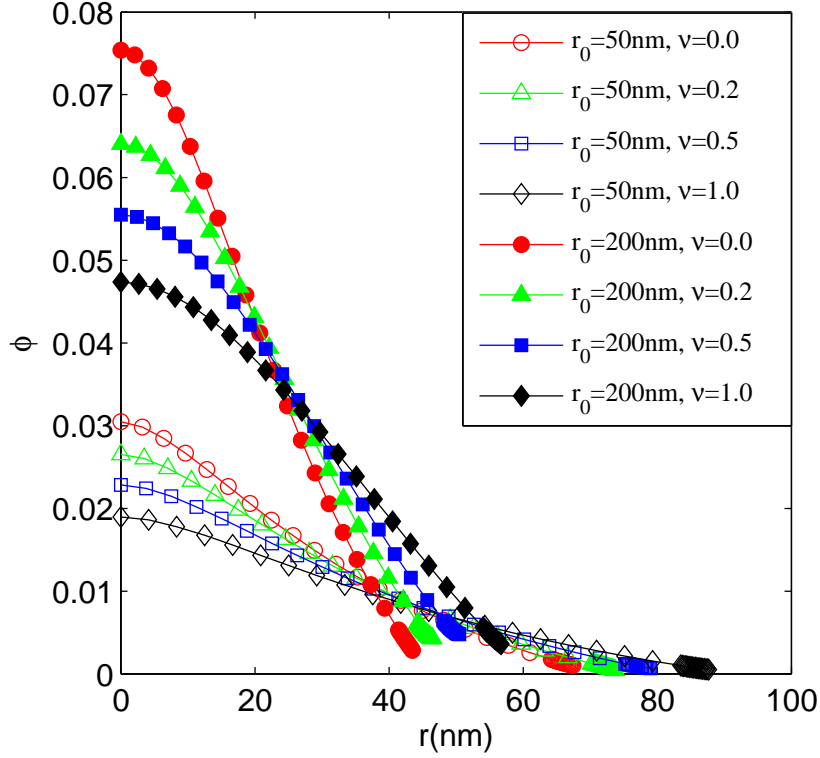}
\caption{(Color online) Monomer distribution profiles for $\nu =0.0,0.2,0.5,1.0$ at $r_0=50nm, 200nm$. $pH=3, c_\infty=0.01mol/L$. Other parameters are the same as in Fig. \ref{fig:6}}
\label{fig:8}
\end{figure}
\begin{figure}
\includegraphics[width=0.65\textwidth]{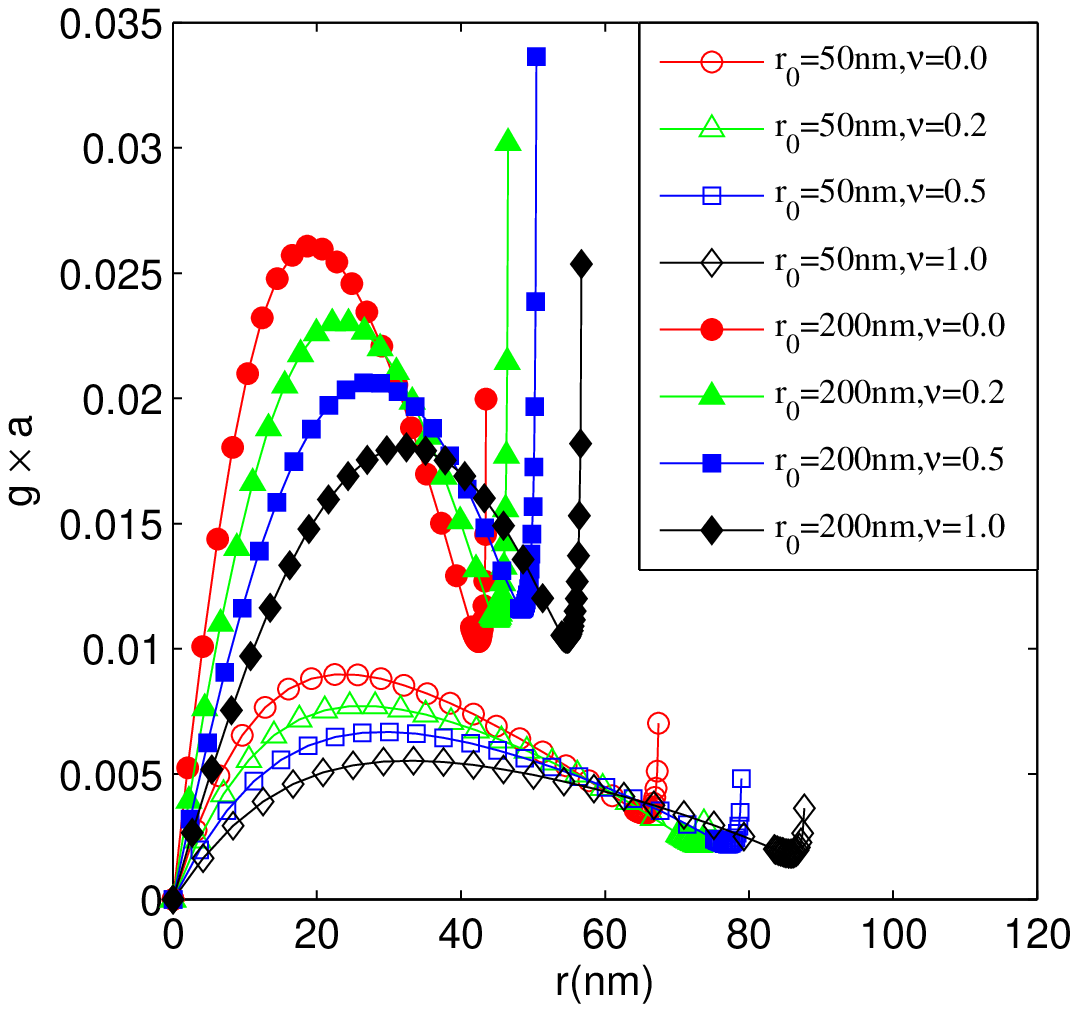}
\caption{(Color online) Chain end distribution profiles for $\nu =0.0,0.2,0.5,1.0$ at $r_0=50nm, 200nm$. $pH=3, c_\infty=0.01mol/L$. Other parameters are the same as in Fig. \ref{fig:6}}
\label{fig:9}
\end{figure}

Fig. \ref{fig:2} shows the polyelectrolyte brush height versus curvature ($1/r_0$) for $pH = 3, c_\infty = 0.1mol/L$; $pH = 3, c_\infty = 0.01mol/L$; $pH = 4, c_\infty = 0.1mol/L$; $pH = 4, c_\infty = 0.01mol/L$.

One of the most important results is that the polyelectrolyte brush height increases with increasing curvature.
That is, the smaller the radius of inner core particle, the larger the polyelectrolyte brush height.
The polyelectrolyte brush height is mainly determined by balancing the elastic forces of the polyelectrolyte chains, monomer excluded volume interactions, electrostatic repulsion between charge monomers, and electrostatic attraction between counterions and polyelectrolyte ions.

The electric field strength in a spherical polyelectrolyte brush is caused by the superposition of the electric fields formed by the charge in the polyelectrolyte chains that are stretched to the inner core and by the electrolyte ions around it.
However, due to the spherical symmetry distribution of the spherical polyelectrolyte brushes, the average electric field will also be directed radially, and the smaller the radius of inner core, the smaller the average electric field strength will be.
Hence, the formular for counterion density results in that when the magnitude of electrostatic potential is low, the number density of counterions is small, so that the attractive force that prevents the polyelectrolyte chain from expanding can be reduced. Finally, in the case, the polyelectrolyte chain can be extended farther.

Fig. \ref{fig:2} also shows that this behavior is stronger at higher pH and lower bulk ionic concentration.
The higher the pH, the lower the hydrogen ion concentration, the more dissociated the polyelectrolytes that produce hydrogen ions by the chemical equilibrium principle.
As a result, the charge of the polyelectrolyte chain is increased, and the electrostatic repulsion is enhanced and finally the brush is expanded.
In addition, increasing the bulk ionic concentration reduces the height of the electric double layer, and the shielding effect occurs at shorter distances, resulting in a weakening of the shielding between the polyelectrolyte chains and a decrease in the chain length.

Fig. \ref{fig:3} shows that the smaller the radius of inner core, the smaller the magnitude of the electric potential inside the brush, and the slower the overall decrease.
This can be explained by the fact that the electric field strength increases with the radius of inner core, as explained previously in Fig. \ref{fig:2}, so the magnitude of the electric field potential increases.
Also, as shown in Fig. \ref{fig:3}, it can be seen that the larger the pH, the smaller the concentration, the larger the magnitude of the electric field potential in the brush region.
As pH increases, the hydrogen ion concentration decreases, resulting in a strong decomposition reaction of polyelectrolyte ions, which leads to an increase in the charge of the polyelectrolyte chain, resulting in an increase in the magnitude of the electric field strength and an increase in the magnitude of the electric field potential.

On the other hand, when the bulk ionic concentration decreases, the shielding effect of the electrolyte solution is weakened, and the electrostatic repulsive interaction between the polyelectrolyte monomers is enhanced, resulting in an increase in the electric field strength and finally an increase in the magnitude of the electric field potential. 

Fig. \ref{fig:4} displays monomer distribution profiles for different pH, bulk ionic concentration and radius of the inner core $r_0=50nm,100nm,150nm, 200nm$.

Fig. \ref{fig:4} shows the monomer distribution function inside the polyelectrolyte brush layer for different radii of inner core at different pH and concentrations. Fig. \ref{fig:4} shows that the smaller the radius of inner core, the smaller the monomer distribution function near the charged wall, and the faster and slower the decreasing trend. This is due to the fact that the smaller the radius of inner core, the longer the polyelectrolyte chain, and therefore, on average, the lower the monomer density.

Fig. \ref{fig:4} represents that the higher the pH, the lower the electrolyte ion concentration, the lower the monomer density.
This can be seen by considering the fact that the higher the pH, the higher the electrolyte concentration, the longer the polyelectrolyte brush height, and the normalization condition of the monomer density, as shown in Fig. \ref{fig:2}

Fig. \ref{fig:5} displays the chain end distribution function inside the polyelectrolyte brush layer with position for various radii of inner core at different pH and concentrations.

Fig. \ref{fig:5} shows that a smaller radius of the inner core gives a smaller end chain density within the polyelectrolyte brush layer. This is due to two reasons.
First, the smaller the radius of inner core, the longer the polyelectrolyte brush layer, and the lower the average density of the end chain distribution. The curvature effect is added to the normalization condition, thereby reducing the chain end distribution over the entire range.
This is attributed to the role of the term $r^2/{r_0}^2$ in the normalization condition of the chain end distribution.

Fig. \ref{fig:5} represents that the larger pH and the smaller the bulk ionic concentration, the smaller the polyelectrolyte chain end distribution.
This can be seen by considering the fact that the higher pH and the higher the bulk ionic concentration, the longer the polyelectrolyte brush height, and the chain end distribution normalization condition, as shown in Fig. \ref{fig:2}

Fig. \ref{fig:6}(a) depictsthe polyelectrolyte brush height for various radius of the inner core at $pH=3, c_\infty=0.01mol/L$. We can see that the stronger the monomer excluded volume effect, the greater the height of the polyelectrolyte brush layer under any radius.
Fig. \ref{fig:6}(b) displays the polyelectrolyte brush height for various $pH$ and bulk ionic concentrations. It is shown that increasing the monomer excluded volume effect yields an increase in the height of the polyelectrolyte brush under any $pH$ and bulk ionic concentration.

Fig. \ref{fig:7} shows the electrostatic potential distribution inside the polyelectrolyte brush layer for $\nu = 0.0, 0.2, 0.5, 1.0$, $r_0 = 50nm, 200 nm$.

Fig. \ref{fig:7} represents that the larger the monomer excluded volume effect, the smaller the magnitude of the electrostatic potential inside the polyelectrolyte layer, and that such behavior is weakened with the radius of inner core.
First, the stronger the monomer excluded volume effect, the slower the electric field magnitude will eventually change as the polyelectrolyte brush layer height increases, and the magnitude of the electrostatic potential will decrease. On the other hand, the larger the radius of the inner core decreases, the larger the increasing trend will be, as shown in Fig. \ref{fig:6}.

Fig. \ref{fig:8} depicts the monomer distribution inside the polyelectrolyte brush layer for $pH = 3, c_\infty = 0.01 mol/L$, $\nu = 0.0, 0.2, 0.5, 1.0$, $r_0 = 50nm, 200 nm$.

Fig. \ref{fig:8} denotes that the larger the monomer excluded volume effect, the smaller the monomer density near the inner core surface, and that such behavior is enhanced with radius of inner core.
   First, the stronger the monomer excluded volume effect, the larger the height of the polyelectrolyte brush layer, the lower the monomer density. On the other hand, an decrease in radius of inner core provides an increase in the height of polyelectrolyte layer, consequently, the monomer density distribution becomes smaller. Therefore, in the case of the smaller radius, the diffrerence in the monomer density between the cases having different values of excluded volume interaction becomes smaller. 

Fig. \ref{fig:9} displays the chain end distribution inside the polyelectrolyte brush layer for $pH = 3, c_\infty = 0.01 mol/L, \nu = 0.0, 0.2, 0.5, 1.0, r_0 = 50nm, 200 nm$.

Fig. \ref{fig:9} means that the larger the monomer excluded volume effect, the smaller the chain end distribution near the inner core surface, and that such behavior is enhanced with the radius of inner core.
This is explained by the fact that the stronger the monomer excluded volume effect, the larger the polyelectrolyte brush layer height, and the smaller the chain end distribution. 

\begin{figure}
\includegraphics[width=\linewidth]{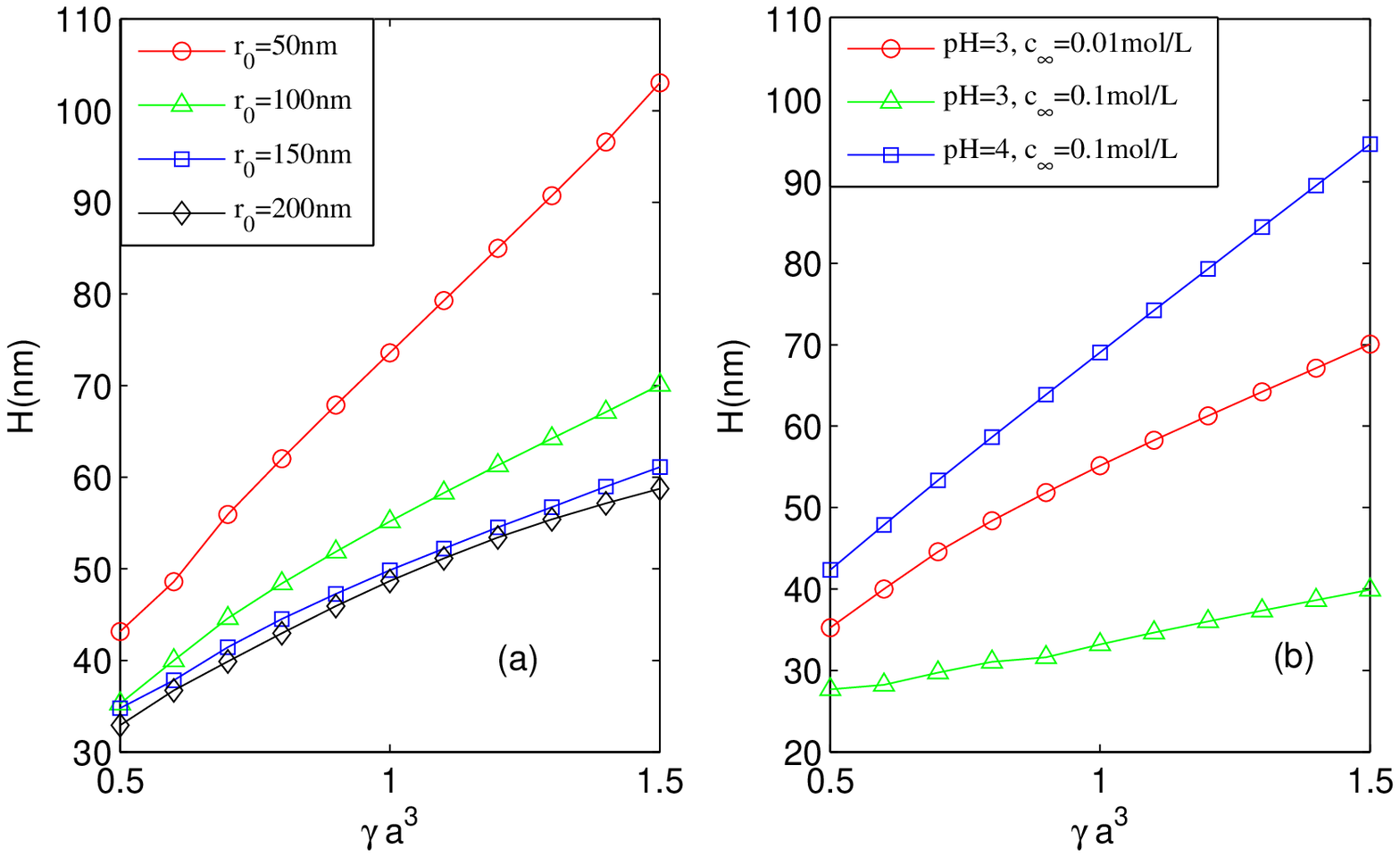}
\caption{(Color online) Variation of equilibrium brush height $H$ with $\gamma a^3$  (a) for $r_0=50nm,100nm,150nm,200nm$ at $pH=3, c_\infty=0.01mol/L$; (b) for $pH=3, 4; c_\infty=0.01mol/L, 0.1mol/L$ at $r_0=100nm$. Other parameters are the same as in Fig. \ref{fig:2}}
\label{fig:10}
\end{figure}

Fig. \ref{fig:10} displays the polyelectrolyte brush height versus $\gamma a^3$ for $pH= 4 $ and $ c_\infty = 0.01 mol/L$, $\nu = 0.2, r_0 = 50nm, 100nm, 150nm, 200 nm$.

From \ref{fig:10}(a), it can be seen that for any radius of inner core, the height of polyelectrolyte brushes increase with the increase in the monomer density of polyelectrolyte brushes.
Furthermore, in the case of increasing $\gamma a^3$, the smaller the radius, the stronger the brush height increases, and the difference in brush height between the cases with different radii of inner core becomes larger.
This is due to the fact that when the charged node density of the chains increases linearly, the electrostatic attraction that prevents the polyelectrolyte chain from swelling due to the curvature effect decreases more rapidly due to the curvature effect term with the charged node density, thus weakening the force that hinders the chain expansion.

Fig. \ref{fig:10}(b) depicts the polyelectrolyte brush height under different pH and concentration conditions.
As shown in \ref{fig:10}(b), at all pH and concentration conditions, the polyelectrolyte brush height decreases with increasing polyelectrolyte chain end distribution.

\begin{figure}
\includegraphics[width=\linewidth]{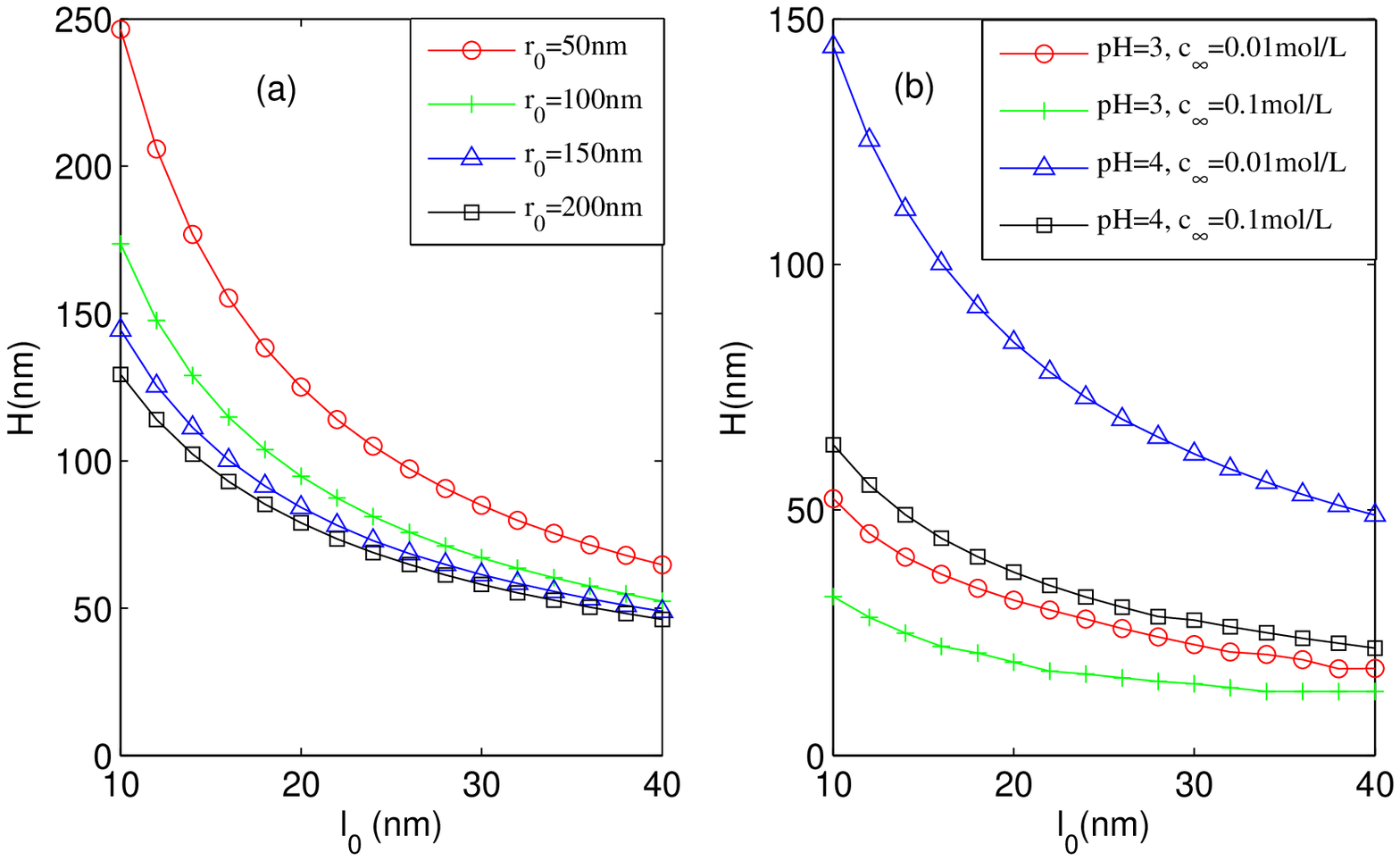}
\caption{(Color online) Variation of equilibrium brush height H with $l_0$  (a) for  $r_0=50nm, 100nm, 150nm, 200nm$ at $pH=4, c_\infty=0.01mol/L$; (b) for $pH=3, 4; c_\infty=0.01mol/L,0.1mol/L$ at $r_0=150nm$. Other parameters are the same as in Fig. \ref{fig:2}}
\label{fig:11}
\end{figure}

Fig. \ref{fig:11} shows the polyelectrolyte brush height versus $l_0$ for $pH=4, c_\infty = 0.01 mol/L$, $\nu = 0.2, r_0 = 50nm, 100nm, 150nm$ and $200 nm$.

From Fig. \ref{fig:11}(a), we can see that the larger the distance between the polyelectrolyte chains, the smaller the height of the polyelectrolyte brush layer.
Furthermore, when increasing lateral separation between chains, the smaller the radius of inner core, the stronger the brush height decreases, and the effect of the difference between the radius of inner core becomes smaller.
This can be understood from the fact that as the lateral separation between the chains increases, the electrostatic repulsive force between the charged monomers in the polyelectrolyte brushes decreases and the brush height decreases.

Fig. \ref{fig:11}(b) shows the polyelectrolyte brush height under different pH and concentration.
As shown in Fig. \ref{fig:11}(b), the polyelectrolyte brush height decreases with increasing the lateral separation between polyelectrolyte chains at all pH and concentration conditions. 

We believe that the present study will potentially change the view of metal nanoparticles grafted with pH-responsive polyelectrolytes, providing new insights into many branches of nanotechnology such as drug delivery system and emulsion stabilization as well as functionalized surface manipulation. \cite%
{murdoch_macromol_2016,cheesman_lang_2013,willot_lang_2015,murdoch_macromol_2018,hariharan_macromol_1998,szleifer_lang_2011,szleifer_sm_2017,szleifer_jpcc_2016,seidel_macromol_2005}

To improve the present theory, we can account for ionic size in electrolyte solutions \cite%
{bikerman_1942, andelman_1997, iglic_1996, bukov_jsm_2022}. However, since we consider the cases where salt concentration and polyelelctrolyte charge density are not high, excluded volume effect of electrolyte ions is negligible, as mentioned in \cite%
{sin_jcp_2022}. 
 In the future, we will try to obtain a more complete model of the pH-responsive polyelectrolyte brushes considering other effects such as  solvent polarization  \cite%
{ iglic_2010, iglic_2012, iglic_2015, sin_2015, sin_2016_1, sin_2016, das_pre_2014, sin_csa_2017}, ionic correlation  \cite%
{ bazant_2011} and the polarization of ions \cite%
{bukov_jpcc_2021,bukov_EPL_2015, bukov_EA_2018}.

\section{Conclusions}
We investigated the electrostatic and structural properties of a spherical polyelectrolyte brush using a strong stretching theory that takes into account the monomer excluded volume effect and the generalized mass action law.

We found that the smaller the inner core particle radius, the thicker the polyelectrolyte brush, and furthermore, the smaller the monomer density function.
It was also demonstrated that the larger the monomer excluded volume interaction, the larger the chargeable site density, the smaller the concentration of electrolyte solution, and the higher pH of electrolyte solution, the thicker the polyelectrolyte brush.

\section{Conflicts of interest}
There are no conflicts to declare.

\section{\bf Data Availavility}
The data that support the findings of this study are available from the corresponding author upon reasonable request.

\end{document}